\documentclass{aastex}
\usepackage{spr-astr-addons}

\RequirePackage{color}

\newcommand{\ha}{H$\alpha$}
\newcommand{\hb}{H$\beta$} 
\newcommand{\hg}{H$\gamma$} 
\newcommand{\kms}{\,km\,s$^{-1}$}

\begin{document}

\title{A collimated wind interpretation \\for the spectral variability of Z~And\\ during its major 2006 eruption}
\shorttitle{The spectral variability of Z~And in 2006}
\shortauthors{Tomov et al.}

\author{N.~Tomov\altaffilmark{1}} \and \author{M.~Tomova\altaffilmark{1}}
\affil{Institute of Astronomy and NAO, Bulgarian Academy of 
           Sciences, P.O. Box 136, 4700 Smolyan, Bulgaria}
\and
\author{D.~Bisikalo\altaffilmark{2}}
\affil{Institute of Astronomy of the Russian Academy of Sciences, 
           48 Pyatnitskaya Str., 119017 Moscow, Russia}


\begin{abstract}
High-resolution observations in the region, centered at $\lambda$\,4400 \AA\, and those of the lines \mbox{He\,{\sc ii}} $\lambda$\,4686, \hb\ and \mbox{He\,{\sc i}} $\lambda$\,6678 of the spectrum of the symbiotic binary Z~And were performed during its outburst in 2006. The line \hb\ had additional satellite high-velocity emission components situated on either side of its central peak. The lines of neutral helium presented two components, consisting of a nebular emission situated close to the reference wavelength and a highly variable P~Cyg absorption. Close to the optical maximum the line \mbox{He\,{\sc ii}} $\lambda$\,4686 was weak emission feature, but with the fading of the light it changed into an intensive emission consisting of a central narrow component and a broad component with a low intensity. The lines of \mbox{N\,{\sc iii}} and \mbox{C\,{\sc iii}} were very broadened. We demonstrate that all of these groups of lines with very different profiles can be interpreted in the light of the same model, where a disc-shaped material surrounding the compact object collimates its stellar wind and gives rise to bipolar outflow.
\end{abstract}


\section{Introduction}

The interacting binary Z~And is a prototype of the classical symbiotic stars and the whole class of symbiotics and consists of a normal M4.5 giant \citep{MS}, very hot compact component and an extended surrounding nebula. After the year 2000 it underwent a long active phase, lasting thirteen years and involving seven optical eruptions. The 2006 eruption is remarkable with the greatest variety of the line profiles of the system indicating a complex flow structure in its nebula.
At that time the line \ha\, had  additional emission satellites with velocities of 1100--1400\kms, located on its either side and indicating bipolar collimated outflow. The line \hg\, had multicomponent P~Cyg absorption reaching velocity of up to about 1500\kms. After the end of July this line had absorption P~Cyg component with lower velocity \citep{BL,sk+09,TTB12}. This behaviour resembles that of the \mbox{He\,{\sc i}} triplet lines during the 2000 outburst \citep{sk+06,TTB08}. 

Collimated jets from astrophysical objects represent very exciting phenomena.
According to the modern theory, their existence is provided by a magnetic accretion disc whose field transforms the potential energy of the accreting material into kinetic energy of the outflowing gas \citep[][and references therein]{Zanni+07,Livio11}. According to the theory, the symbiotic stars with a magnetic disc can have collimated outflow in both of their states -- quiescent and active ones. Collimated outflow, however, can be produced by the stellar wind of the outbursting compact object too, if it is surrounded by an extended disc-shaped structure which provides a small opening angle of the outflowing gas \citep{TTB14}.
 During the 2006 brightening of Z~And we observed on one hand high velocity satellite emission components of the line \hb, indicating bipolar collimated outflow and, on the other hand, multicomponent P~Cyg absorption of the helium lines with velocities in a broad range -- from 100~km\,s$^{-1}$, to about 1\,500~km\,s$^{-1}$, reaching thus the velocity of the satellite emission. These data propose that the collimated outflow and the P~Cyg wind are probably connected and can be related to the same flow.  

The mean aim of this work is to show that spectral lines of Z~And with very different behaviour during its major 2006 eruption can be considered in the light of the same model -- the model of the collimated stellar wind. Another aim is to obtain the mass-loss rate of the outbursting component at different times during the eruption.   

\section{Observations and data reduction}

The regions of the lines \mbox{He\,{\sc i}} $\lambda$\,6678, \mbox{He\,{\sc ii}} $\lambda$\,4686, and that centered at $\lambda$\,4400 \AA\, were observed on fourteen nights during 2006 July -- December 
with the Photometrics CCD array mounted on the Coude spectrograph of the 2m Ritchey-Chretien-Coude (RCC) telescope of the Rozhen National Astronomical Observatory.
 In addition the region of the line \hb\ was observed on four nights from September till December 2006 (Table~\ref{journal}). In our previous work \citep{TTB12} we considered the behaviour of the lines \ha\ and \hg\ of Z~And during the same period. Some of the lines, treated in the current work, are in the \ha\ and \hg\ regions. The spectra in the regions of \hb\ and \mbox{He\,{\sc ii}} $\lambda$\,4686 were reduced in the same way as those in the \ha\ and \hg\ regions described in the work of \citet{TTB12}. According to the continuum analysis of \citet{sk+09} the flux at the position of the line \mbox{He\,{\sc i}} $\lambda$\,6678 of Z~And is practically equal to its $R_{\rm C}$\, flux. Then we used the $R_{\rm C}$\, flux to calculate the \mbox{He\,{\sc i}} $\lambda$\,6678 line flux.


\begin{table}[t]
 \caption{List of the observations}
  \label{journal}
  \centering                          
 \begin{tabular}{@{}llcl}
  \hline
    \noalign{\smallskip}  
	Date & JD$-$ & Orb. & Spectral \\
	     & 2\,453\,000 & phase & region \\
    \noalign{\smallskip} 
  \hline
    \noalign{\smallskip}
  Jul 08   & ~~924.56 & 0.837 & \mbox{He\,{\sc i}}, \mbox{He\,{\sc ii}}, $\lambda$\,4400 \\
  Jul 09   & ~~926.45 & 0.840 & \mbox{He\,{\sc i}}, \mbox{He\,{\sc ii}}, $\lambda$\,4400 \\
  Jul 14   & ~~931.48 & 0.846 & \mbox{He\,{\sc i}}, \mbox{He\,{\sc ii}}, $\lambda$\,4400 \\
  Jul 19   & ~~935.59 & 0.852 & \mbox{He\,{\sc i}}, \mbox{He\,{\sc ii}} \\
  Jul 20   & ~~936.58 & 0.853 & $\lambda$\,4400 \\
  Aug 13   & ~~960.51 & 0.885 & \mbox{He\,{\sc ii}}, $\lambda$\,4400 \\
  Sept 07  & ~~985.57 & 0.918 & \mbox{He\,{\sc i}}, \mbox{He\,{\sc ii}}, $\lambda$\,4400 \\
  Sept 08  & ~~986.54 & 0.919 & \mbox{He\,{\sc i}}, \hb, \mbox{He\,{\sc ii}}, $\lambda$\,4400 \\
  Oct 03   & 1012.47 & 0.953 & \mbox{He\,{\sc i}}, \mbox{He\,{\sc ii}}, $\lambda$\,4400 \\
  Oct 04   & 1013.46 & 0.954 & \mbox{He\,{\sc i}}, \hb, \mbox{He\,{\sc ii}}, $\lambda$\,4400\\
  Oct 31   & 1040.37 & 0.990 & \mbox{He\,{\sc i}}, \mbox{He\,{\sc ii}}, $\lambda$\,4400\\
  Dec 01   & 1071.37 & 0.031 & \mbox{He\,{\sc i}}, \hb, \mbox{He\,{\sc ii}}, $\lambda$\,4400 \\
  Dec 02   & 1072.30 & 0.032 & \mbox{He\,{\sc i}}, \hb, \mbox{He\,{\sc ii}}, $\lambda$\,4400 \\
  Dec 30   & 1100.24 & 0.069 & \mbox{He\,{\sc i}}, \mbox{He\,{\sc ii}}, $\lambda$\,4400 \\
    \noalign{\smallskip}
  \hline
 \end{tabular}
\end{table}

We used the ephemeris $\rm {Min(vis)=JD}~2\,442\,666\fd0+758\fd8 \times E$, where the zero phase is at the epoch of the orbital photometric minimum which coincides with that of the spectral conjunction \citep{FL,MK,f+00}.

\section{Analysis and interpretation of the line spectrum}

\subsection{The model}

\begin{figure}[!htb]

	\includegraphics[width=0.47\hsize]{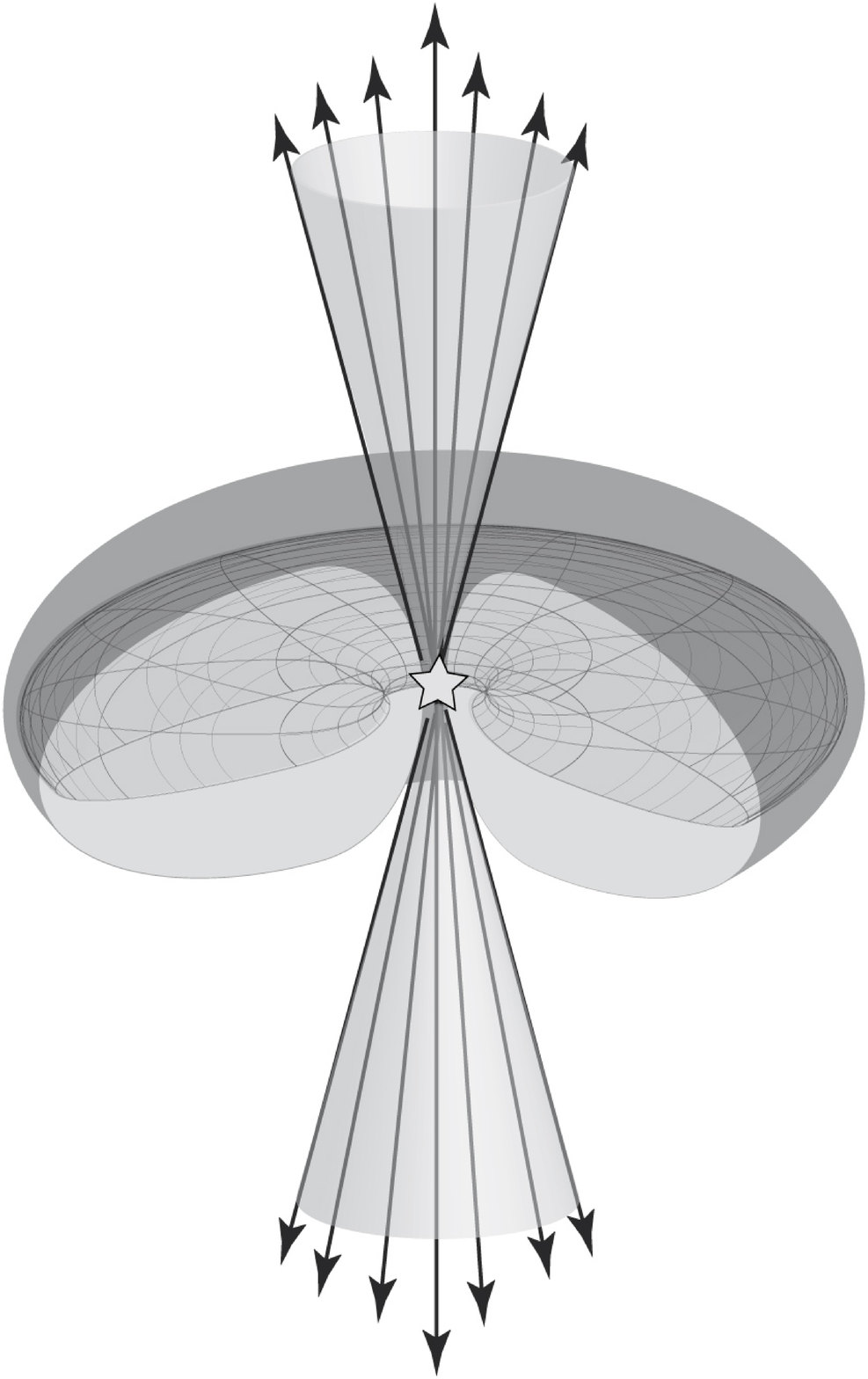}
	\includegraphics[width=0.4\hsize]{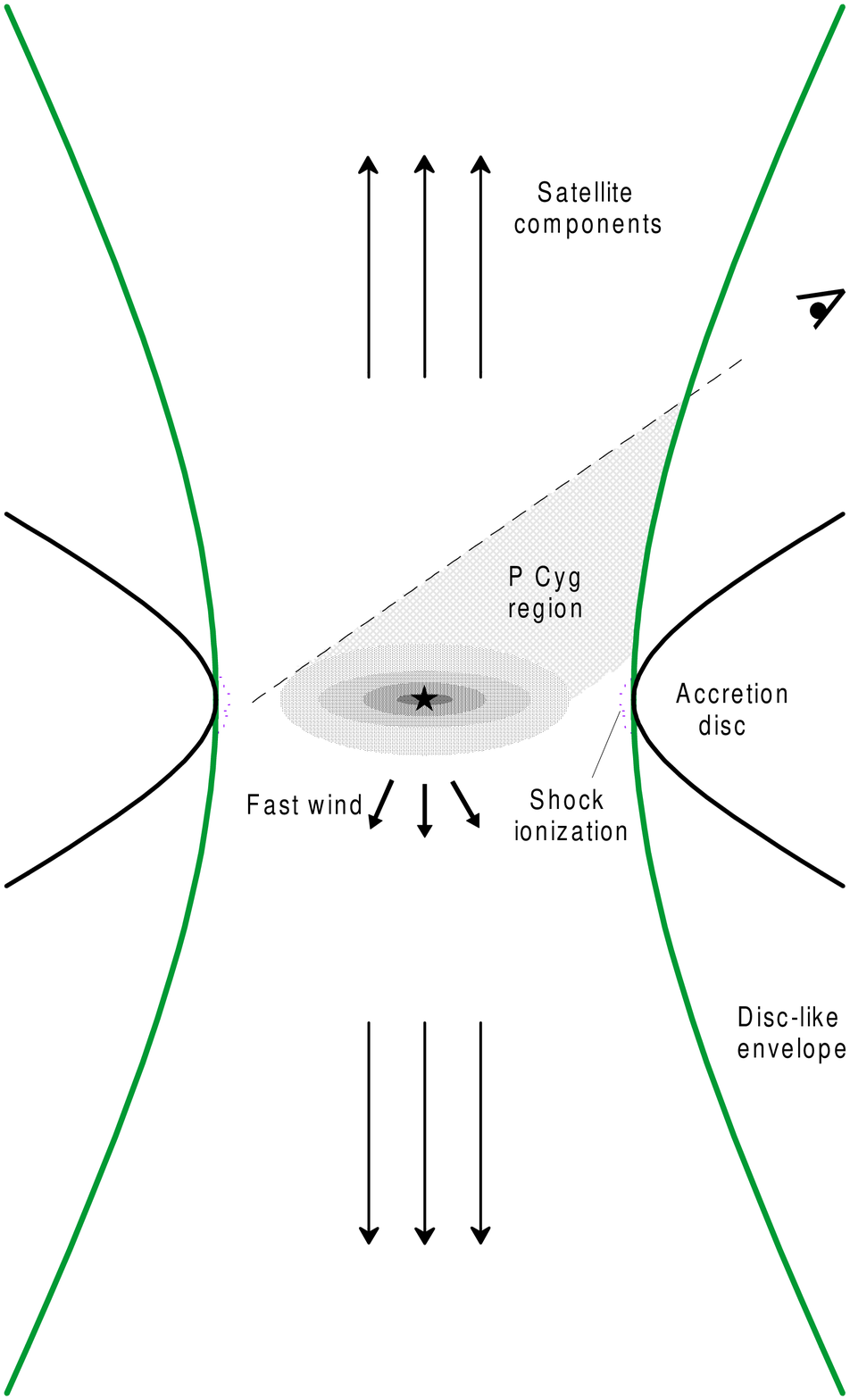}

 \caption{Left panel: Schematic model of the region around the hot component during recurrent strong outburst. Right panel: The same, but in the plane perpendicular to the orbital plane where the emission regions are shown \citep{TTB14}.}
 \label{model2}
\end{figure}

The line spectrum was interpreted in the light of the model proposed by \citet{TTB14}. It is supposed that after some strong outburst of the active phase a part of the ejected material accretes again and as a result of conservation of its initial angular momentum forms an extended envelope covering the initial accretion disc. As a result of existence of centrifugal barrier two hollow cones with a small opening angle appear around the axis of rotation of the envelope. During the next eruption the outflowing material (stellar wind) from the outbursting compact object propagates only through these hollow cones giving rise to high velocity satellite components of the spectral lines (Fig.~\ref{model2}). 

\subsection{H$\beta$ line} 

\begin{figure}[!tH]
    \includegraphics[angle=270,width=\linewidth]{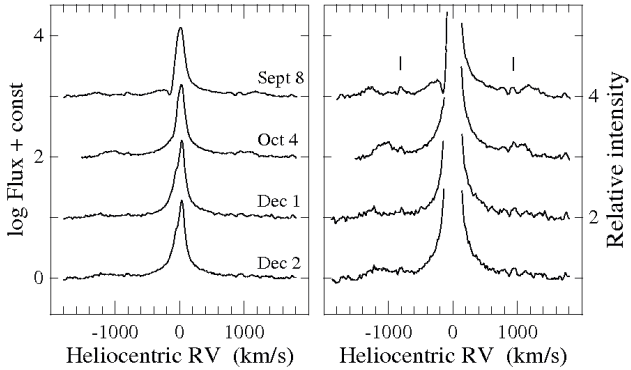}
\caption[Time evolution of the \hb\, line]
				{Time evolution of the \hb\, line: the whole line (left panel), and the lower part of the spectrum (right panel). The emission lines of \mbox{Cr\,{\sc ii}}\,30 $\lambda$\,4848 and $\lambda$\,4876 are marked.}
\label{Hb_06}
\end{figure}

\begin{table*}
 \centering
  \caption{The \hb\, line data.}
    \label{hb_data}
  \begin{tabular}{lrccccccr}
  \hline
    \noalign{\smallskip}
       Date  & $F$(t)\tablenotemark{a} & \multicolumn{3}{c}{Blue} & \multicolumn{3}{c}{Red} & $\dot M_{\rm cw}$\tablenotemark{b} \\
       \cline{3-5} \cline{6-8} \noalign{\smallskip}
     				& & RV & $F$ & $\dot M$ & RV & $F$ & $\dot M$ & \\
    \noalign{\smallskip}
  \hline
     \noalign{\smallskip}
     
Sept 8  & 90.170 & $-$1291 & 1.198 & 0.90 & 1196 & 1.800 & 1.38 & 2.28 \\
 Oct 4  & 91.199 & $-$1074 & 2.507 & 1.79 & 1078 & 1.393 & 0.88 & 2.67 \\
 Dec 1  & 73.248 & $-$1261 & 0.804 & 0.73 & \ldots & \ldots & \ldots & $>$0.73 \\
 Dec 2  & 76.814 & $-$1177 & 1.408 & 1.71 & \ldots & \ldots & \ldots & $>$1.71 \\

    \noalign{\smallskip}
\hline
\end{tabular}

\tablenotetext{a}{$F$(t) is the total line flux.}
\tablenotetext{b}{$\dot M_{\rm cw}$ is a sum of the mass-loss rates based on the satellite components.}
\tablecomments{All fluxes are in units of $10^{-12}$ erg\,cm$^{-2}$\,s$^{-1}$, the mass-loss rate -- in units of $10^{-7}$(d/1.12kpc)$^{3/2}$~M$_{\sun}$\,yr$^{-1}$ and the radial velocity -- in units of km\,s$^{-1}$.}
\end{table*}

Figure~\ref{Hb_06}
shows the evolution of the H$\beta$ line in one short period of time from September to December 2006. At that time the H$\beta$ profile was similar to the profile of H$\alpha$ \citep{TTB12} and consisted of a strong central emission component (core), located around the reference wavelength, broad wings extended to about 1\,500~km\,s$^{-1}$ from the center of the line and additional emission features on both sides of the central component. On the spectrum of Sept. 8 the line had one blueshifted absorption dip in its emission profile at a velocity of about $-150$~km\,s$^{-1}$ from its center. In December the central emission presented a weak shoulder on its short-wavelength side at a velocity of $-70$~km\,s$^{-1}$ from the line center like the absorption dip in the emission profile of H$\alpha$ at that time. The high-velocity satellite emission components were fitted with a Gaussian and the other part of the line (the core together with the wings) -- with two or three Lorentzians. The uncertainty of the equivalent width of the satellite emissions was not more than 30 per cent and that of the whole line -- about 2 per cent. According to the theory the profile of the spectral line determined by radiation damping is approximated with a Lorentzian function. That is why we assume that the broad H$\beta$ wings are determined mainly by radiation damping. It is possible, however, that the stellar wind of the outbursting object giving rise to the broad component of the line \mbox{He\,{\sc ii}} 4686 (Sect. 3.4.1) and the region of shock waves produced by the collision of the wind with the accretion disc/envelope \citep{Drake,Ionov12} contribute to these wings, too. 

The line H$\beta$ presented additional satellite emission components with a velocity of up to about 1\,300~km\,s$^{-1}$ situated on either side of the central component indicating bipolar collimated outflow from the system like the line H$\alpha$. The evolution of the spectrum shows that their flux decreases with time due to decrease of the mass-loss rate of the compact object (Fig.~\ref{Hb_06}, Table~\ref{hb_data}).

\subsection{Helium lines}

\begin{figure}[!pH]
\center
    \includegraphics[width=\linewidth]{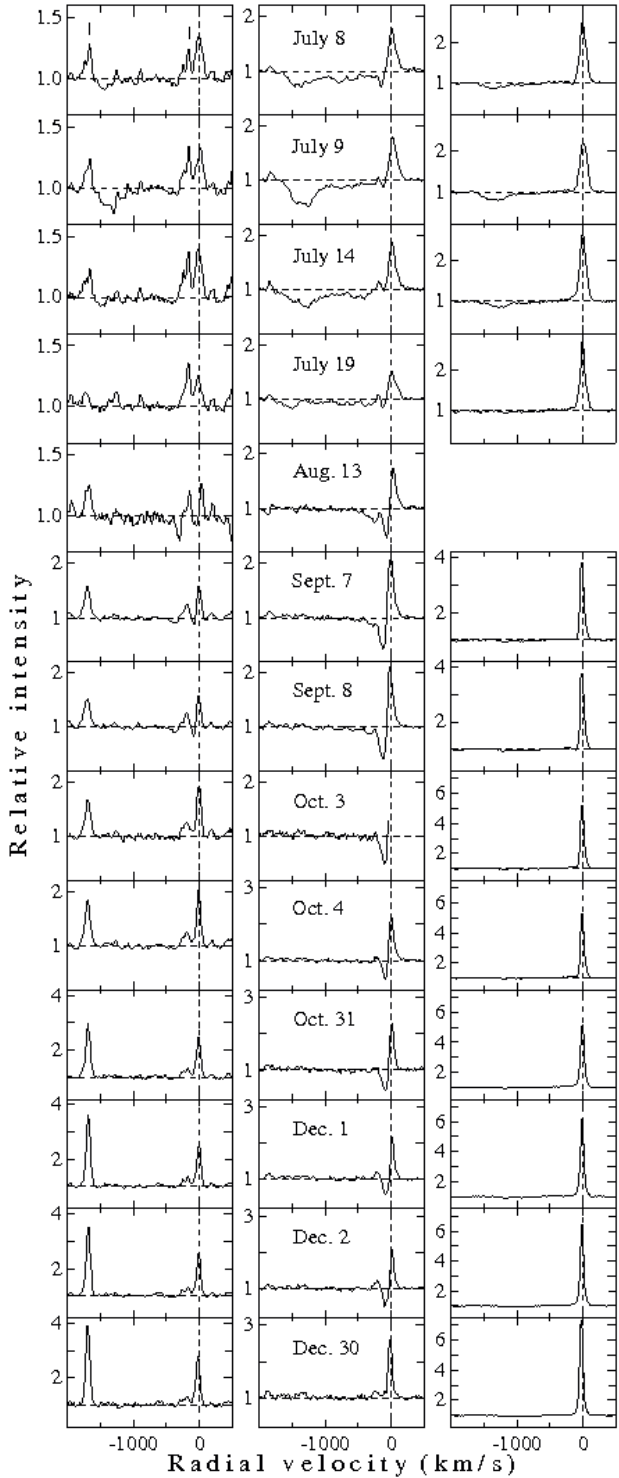}
\caption[]
				{Time evolution of helium lines: \mbox{He\,{\sc i}} $\lambda$\,4388 (left panel), \mbox{He\,{\sc i}} $\lambda$\,4471 (middle panel) and \mbox{He\,{\sc i}} $\lambda$\,6678 (right panel). The vertical dashed line shows the position of the reference wavelength and the horizontal dashed line marks the level of the 
				local continuum. The emission lines \mbox{O\,{\sc iii}} $\lambda$\,4363 and \mbox{Fe\,{\sc ii}} $\lambda$\,4385 are marked on the spectrum of July 8.}
\label{HeI}
\end{figure}

In the period of quiescence prior to the 2000--2013  active phase the lines of \mbox{He\,{\sc i}} had an ordinary nebular profile \citep{TTB08}.

In July 2006 the singlet line \mbox{He\,{\sc i}} $\lambda$\,4388~\AA\ had a high velocity P~Cyg component with the same position, like the absorption component of H$\alpha$ and the most blue shifted component of H$\gamma$ \citep{TTB12}. In July 8--14, however, the helium line had an additional absorption at lower velocities of $-320 \div -520$\kms\, with multicomponent structure (Fig.~\ref{HeI}, left panel). Some part of this absorption is probably not seen because of presence in the spectrum of the emission line \mbox{Fe\,{\sc ii}} $\lambda$\,4385~\AA. On August 13 the line \mbox{He\,{\sc i}} $\lambda$\,4388~\AA\ had a multicomponent absorption occupying an interval $-15 \div -400$\kms\, and the \mbox{Fe\,{\sc ii}} line was thus inside it. The behaviour of the helium line in September was similar to that in August. After that the line was purely in emission.

In July 8--14 the singlet line \mbox{He\,{\sc i}} $\lambda$\,6678~\AA\ had a high velocity P~Cyg component with the same position, like the other singlet line and only on July 8 it had a weak blue absorption with a low velocity of $-170$\kms. On July 19 and September 7 it was purely emission line and had a weak absorption with a low velocity of $-110$\kms\, on September 8 again, which means that it underwent at that time fast variation on a time scale of one day. After that the line \mbox{He\,{\sc i}} $\lambda$\,6678~\AA\ was purely in emission (Fig.~\ref{HeI}, right panel).

\begin{figure}[!tH]
\center
    \includegraphics[width=.48\textwidth]{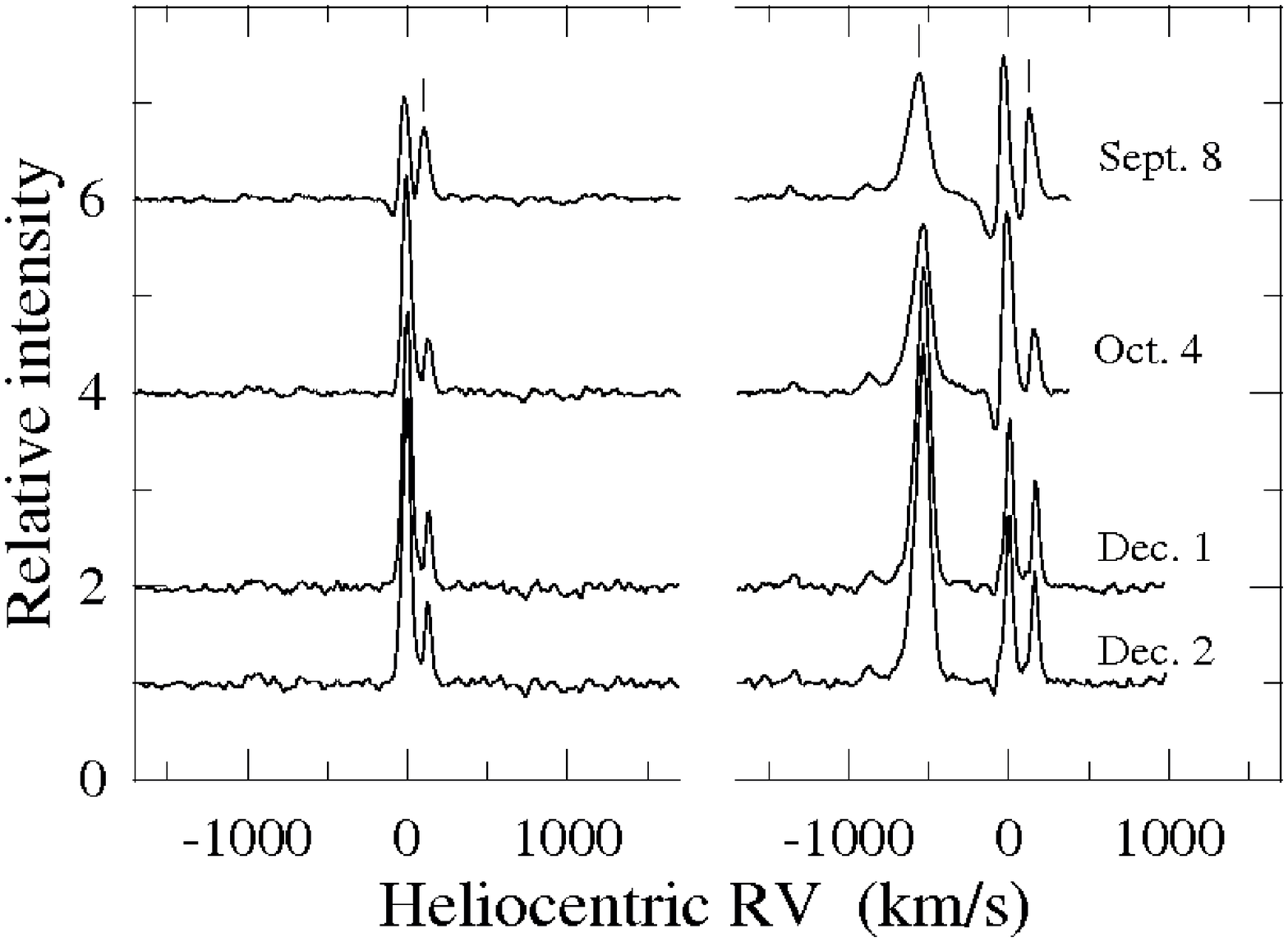}
    \includegraphics[width=.48\textwidth]{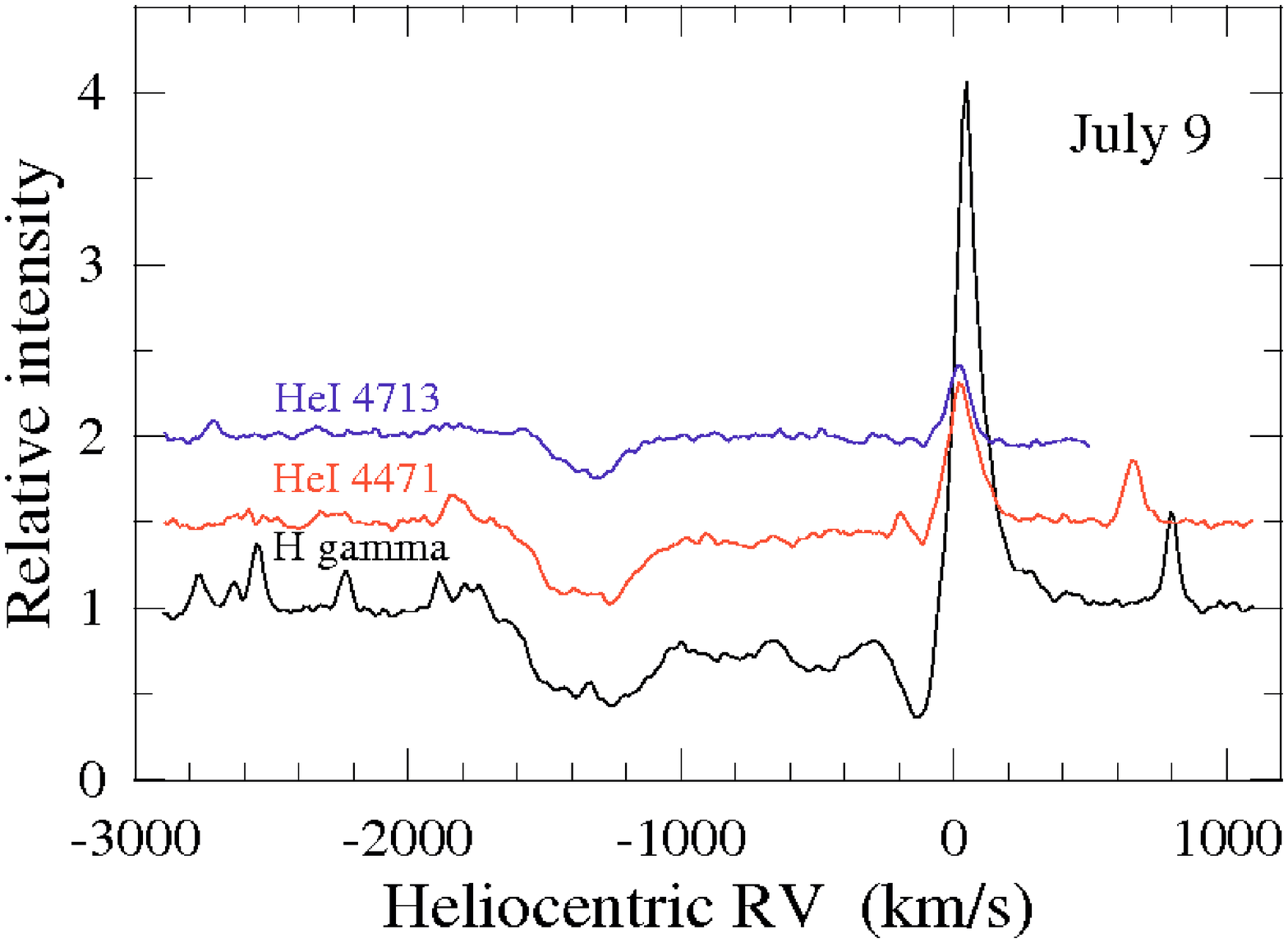}
\caption{
{\it Upper panel:} Time evolution of helium lines: \mbox{He\,{\sc i}} $\lambda$\,4922 (left) and \mbox{He\,{\sc i}} $\lambda$\,5016 (right). The emission lines \mbox{Fe\,{\sc ii}} $\lambda$\,4924, \mbox{O\,{\sc iii}} $\lambda$\,5007 and \mbox{Fe\,{\sc ii}} $\lambda$\,5018 are marked on the spectrum of Sept. 8.
{\it Lower panel:} The profiles of the \hg\ and the \mbox{He\,{\sc i}} 							triplet lines on July 9.
				}
\label{HeI_singl&HeIHg}
\end{figure}

The singlet lines \mbox{He\,{\sc i}} $\lambda$\,4922~\AA\ and $\lambda$\,5016~\AA\ are in the spectral range of H$\beta$ and were observed only in four nights in September -- December. The line \mbox{He\,{\sc i}} $\lambda$\,4922~\AA\ had a P~Cyg absorption with a low velocity of $-100$\kms\, only on September 8 and the line \mbox{He\,{\sc i}} $\lambda$\,5016~\AA\ all the time (Fig.~\ref{HeI_singl&HeIHg}, left panel).

\begin{table*}[tH]
\footnotesize
 \caption{The fluxes of the emission lines of Z~And during its 2006 outburst in units $10^{-12}$ erg\,cm$^{-2}$\,s$^{-1}$}
  \label{emission_lines_2006}
  \centering
 \begin{tabular}{@{}lccccccccccccc@{}}
    \noalign{\smallskip}
  \hline
    \noalign{\smallskip}
Date & \hg\tablenotemark{a} & \mbox{[O\,{\sc iii}]} & \mbox{He\,{\sc i}} & \mbox{He\,{\sc i}} & \mbox{Fe\,{\sc ii}} & \mbox{He\,{\sc i}} & \mbox{He\,{\sc i}} & \mbox{Fe\,{\sc ii}} & \mbox{[O\,{\sc iii}]} & \mbox{[O\,{\sc iii}]} & \mbox{He\,{\sc i}} & \mbox{Fe\,{\sc ii}} & \mbox{He\,{\sc i}} \\
	   & & $\lambda$\,4363 & $\lambda$\,4388 & $\lambda$\,4471 & $\lambda$\,4629 & $\lambda$\,4713 & $\lambda$\,4922 & $\lambda$\,4924 & $\lambda$\,4959 & $\lambda$\,5007 & $\lambda$\,5016 & $\lambda$\,5018 & $\lambda$\,6678 \\	
    \noalign{\smallskip}
  \hline
    \noalign{\smallskip}
Jul 8 & 26.697 & 2.047 & 3.547 & 7.421 & 2.828 & 3.839 &  &  &  &  &  &  & ~~7.525 \\
Jul 9 & 28.791 & 2.173 & 3.311 & 7.946 & 2.768 & 2.846 &  &  &  &  &  &  & ~~7.231 \\
Jul 14& 35.054 & 1.902 & 3.566 & 8.098 & 3.459 & 3.099 &  &  &  &  &  &  & ~~9.568 \\
Jul 19& & & & & 2.898 & 1.798 &  &  &  &  &  &  & ~~7.974 \\
Jul 20& 45.003 & 1.187 & 2.477 & 6.289 & & &  &  &  &  &  &  & \\
Aug 13& 27.873 & 1.574 & 0.794 & 3.584 & 1.781 & 1.128 &  &  &  &  &  &  & \\
Sep 7 & 32.064 & 3.303 & 2.293 & 5.786 & 2.045 & 2.543 &  &  &  &  &  &  & ~~7.101 \\
Sep 8 & 29.453 & 3.158 & 2.094 & 4.911 & 2.055 & 2.314 & 3.122 & 2.492 & 3.125 & ~~9.783 & 4.183 & 2.836 & ~~6.844 \\
Oct 3 & 27.676 & 3.497 & 3.660 & & 1.970 & 2.825 &  &  &  &  &  &  & ~~9.852 \\
Oct 4 & 30.550 & 3.901 & 3.741 & 4.662 & 2.000 & 2.591 & 6.845 & 1.563 & 3.586 & 11.497 & 5.859 & 1.927 & ~~9.630 \\
Oct 31& & 6.714 & 4.558 & 3.566 & 1.977 & 2.269 &  &  &  &  &  &  & ~~9.658 \\
Dec 1 & & 7.017 & 3.914 & 2.412 & 1.828 & 1.276 & 6.374 & 1.367 & 3.734 & 11.758 & 3.294 & 1.896 & 10.053 \\
Dec 2 & & 7.027 & 4.062 & 2.344 & 1.797 & 1.481 & 6.504 & 1.493 & 4.043 & 12.373 & 3.279 & 2.107 & 10.228 \\
Dec 30& & 6.943 & 3.981 & 3.646 & 1.883 & 2.414 &  &  &  &  &  &  & 10.785 \\
    \noalign{\smallskip}
  \hline
    \noalign{\smallskip}
 \end{tabular}
\tablenotetext{a}{The \hg\ flux is for the whole observed emission including the red wing of the broad component. The rest of the \hg\ data together with the full analysis of its profile are presented in the work of \citet{TTB12}.}
\end{table*}

Prior to July 19 the triplet line \mbox{He\,{\sc i}} $\lambda$\,4713~\AA\, (Fig.~\ref{HeII_06}) had a high velocity P~Cyg component with the same position like the absorption components of the Balmer and singlet lines and only on July 8 this line had in addition a weak low velocity blue absorption at a position of about $-150$\kms. On August 13 this line had a multicomponent P~Cyg absorption showing velocities of the outflowing material in the interval 60$\div$350\kms. The P~Cyg absorption reached its maximal depth in September -- October and the velocity of the most intensive component was of about $-100$\kms. 

In July a blue absorption with a high velocity of $-1300 \div -1400$\kms\, was observed in the triplet line \mbox{He\,{\sc i}} $\lambda$\,4471~\AA\ too, but this line had also additional absorption components, occupying an interval of lower velocities like the line H$\gamma$ (Figs.~\ref{HeI}, middle panel and \ref{HeI_singl&HeIHg}, right panel). In August and September the helium line had a multicomponent P~Cyg absorption showing velocities of the outflowing material in the interval $20 \div 500$\kms. The P~Cyg absorption reached its maximal depth in September. Since October the triplet lines have only a low velocity P~Cyg absorption which was observed up to the beginning of December (Figs.~\ref{HeI} and \ref{HeII_06}). The line fluxes of the helium lines are listed in Table~\ref{emission_lines_2006},
where data of some other lines placed in our observed regions are included.

In the framework of our model (Fig.~\ref{model2}) the P~Cyg components of the \mbox{He\,{\sc i}} lines can be interpreted as absorption by the outflowing material which is projecting onto the observed photosphere of the outbursting compact object (the flat shell). The absence of high velocity emission components of the \mbox{He\,{\sc i}} lines suggests that the emission region of \mbox{He\,{\sc i}} was small.

A main characteristics of the behaviour of the helium lines is transformation of their broad multicomponent P~Cyg absorption into a low velocity P~Cyg feature with the fading of the light after its maximum. The velocity of the outflowing material from the compact object is the highest close to the axis of the cone and the lowest -- close to the ``wall'' of the extended disc-like envelope where its density is the highest (Tomov et al. 2014). When the mass-loss rate of the compact object decreases, it is possible that only the most dense layer placed close to the ``wall'' absorbs the photospheric radiation and in this way only the P~Cyg component at the lowest velocity remains in the line profile. So we could understand the presence only of the low velocity P~Cyg absorption at the end of the collimated ejection episode.

\begin{table*}[!tH]
 \centering
 \caption{The \mbox{He\,{\sc ii}} $\lambda$\,4686 line data. $\dot M_{\rm w}$ is based on both lines \hg\, \citep{TTB12} and \mbox{He\,{\sc ii}} $\lambda$\,4686.}
  \label{heii_data}
 \begin{tabular}{@{}lrcrcccccr@{}}
  \hline 
    \noalign{\smallskip}
	Date & RV(N) & FWHM(N) & $F$(N) & FWHM(B) & FWZI(B) & $\upsilon_{\rm w}$ & $F$(B) & $\dot M$ & $\dot M_{\rm w}$ \\
	     & (km\,s$^{-1}$) & (km\,s$^{-1}$) & & (km\,s$^{-1}$) & (km\,s$^{-1}$) & (km\,s$^{-1}$) & & & \\
    \noalign{\smallskip}
  \hline
    \noalign{\smallskip}

Jul 8  & $<-$200 & \ldots & ~~5.046 & & & & & & \\
Jul 9  &   $-$84 & 213 & ~~0.986 & & & & & & \\
Jul 14 & $<-$200 & \ldots & ~~5.230 & & & & & & \\
Jul 19 &  $-$100 & 219 & ~~2.549 & & & & & & \\
Aug 13 &   $-$89 & 197 & ~~3.540 & & & & & & \\
Sep 7  &   $-$87 & 141 & ~~9.083 & & & & & & \\
Sep 8  &  $-$132 & 265 & ~~5.867 & & & & & & \\
Oct 3  &   $-$58 & 208 & ~~9.552 & & & & & & \\
Oct 4  & $-$57 & 134 & ~~7.580 & 558$\pm$~~40 & 1228 & 600 & 4.945 & 1.14 & 1.14 \\
Oct 31 & $-$24 & 147 & 27.424  & 613$\pm$100  & 1215 & 600 & 3.824 & 0.80 & 1.05 \\
Dec 1  & $-$25 & 160 & 28.249  & 682$\pm$~~40 & 1392 & 700 & 5.330 & 1.02 & 1.14 \\
Dec 2  & $-$25 & 160 & 26.010  & 590$\pm$~~30 & 1240 & 600 & 4.776 & 0.82 & 1.20 \\
Dec 30 & $-$24 & 137 & 42.910  & 634$\pm$~~50 & 1302 & 650 & 4.465 & 0.82 & 1.18 \\

    \noalign{\smallskip} 
  \hline
 \end{tabular}
\tablecomments{N denotes narrow component and B -- broad component. $F$ is in units of $10^{-12}$ erg\,cm$^{-2}$\,s$^{-1}$ and 
$\dot M$ -- in units of 10$^{-7}$(d/1.12~kpc)$^{3/2}$~M$_{\sun}$\,yr$^{-1}$.}
\end{table*}

\subsection{Lines of elements with high ionization degree}

\subsubsection{\mbox{He\,{\sc ii}} $\lambda$\,4686 line}

\begin{figure*}[!tH]
\center
    \includegraphics[width=.9\textwidth]
    {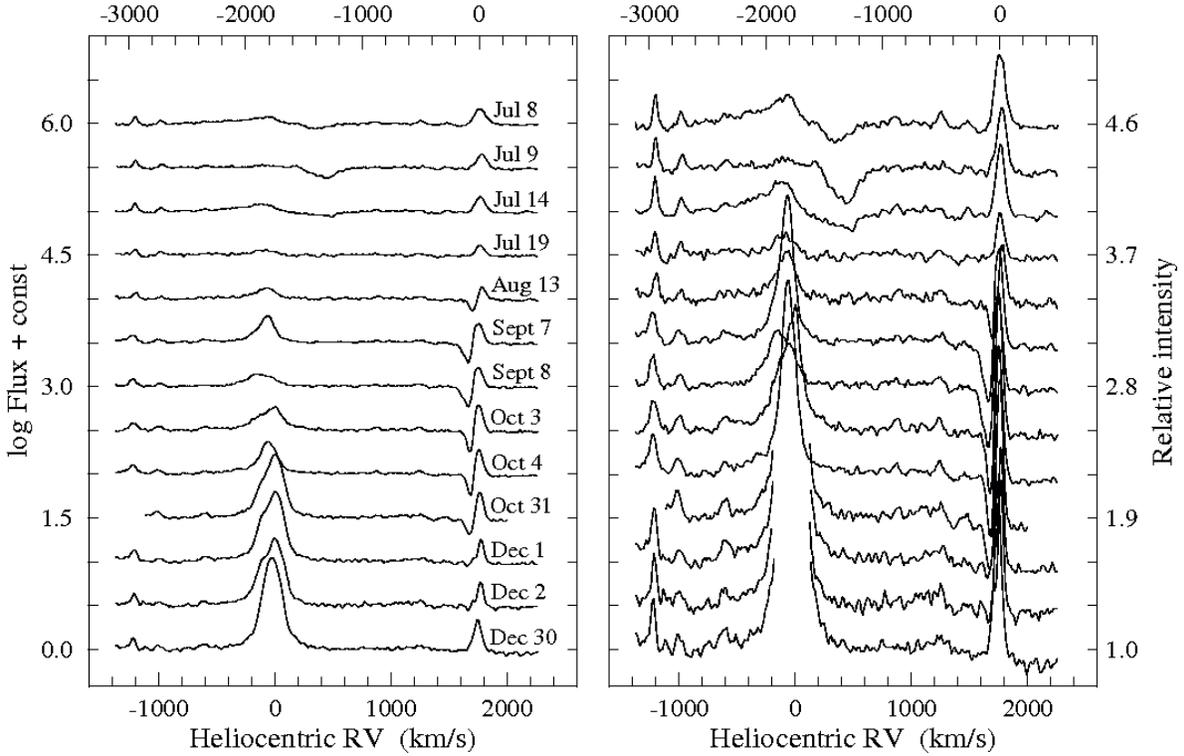}
\caption[The evolution of the profile of the lines \mbox{He\,{\sc ii}} 4686 and \mbox{He\,{\sc i}} 4713]
				{The evolution of the profile of the lines \mbox{He\,{\sc ii}} 4686 and \mbox{He\,{\sc i}} 4713: the whole lines (left panel) and the lower part of the spectrum (right panel) where the \mbox{He\,{\sc ii}} profile in July and August, the broad component of this line and the P~Cyg absorption of the \mbox{He\,{\sc i}} line are better seen. The upper X-axis is about the \mbox{He\,{\sc i}} 4713 line.
				}
\label{HeII_06}
\end{figure*}

\begin{figure}[!tH]
    \includegraphics[width=\linewidth]{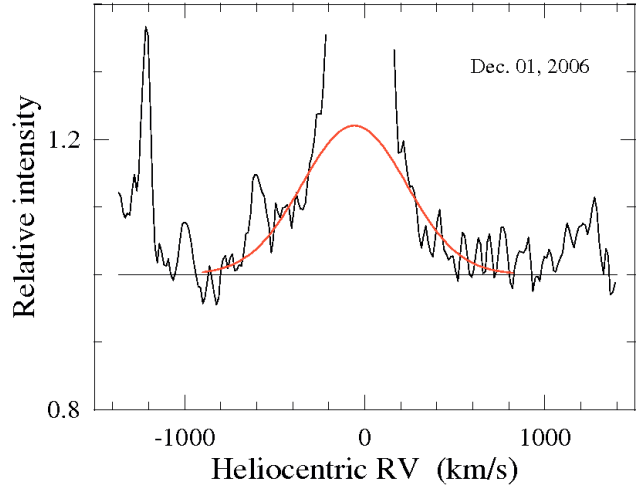}
\caption{The area of the wings of the \mbox{He\,{\sc ii}} 4686 line where the Gaussian approximation is seen.}
\label{he01dec}
\end{figure}

The evolution of the line \mbox{He\,{\sc ii}} 4686 during the period of our observations is shown in Fig.~\ref{HeII_06}. In October -- December it had a two-component profile, consisting of a central narrow emission and a low intensity broad component with FWZI of 1200 km/s like during the previous brightening of Z~And in 2000--2002 \citep{TTB08}. The two 
components had different behaviour with the fading of the light: the energy flux of the narrow one increased whereas the flux of the broad component did not (Table~\ref{heii_data}). Before October this component was hardly visible due to the intensive optical continuum of the system, absorption by the high-velocity P~Cyg component of the line \mbox{He\,{\sc i}} 4713 and its natural variability. At that time it was associated with an emission, exceeding the level of the local continuum, which can not be measured with adequate accuracy, like in the quiescent state of the system \citep{TTB08}. We analysed the broad component by approximating with a Gaussian function (Fig.~\ref{he01dec}) and its parameters obtained with this procedure are listed in Table 5. The error of the equivalent width due to the uncertainty of the continuum level is 8--14 per cent. The broad component 
showed a velocity close to that of the line \hg\ and based on the \hg\ analysis \citep{TTB14} we supposed this component is emitted by the same regions in the system which give rise to the \hg\ broad component, mainly by the stellar wind of the outbursting object. 

In July, at the time of the light maximum a weak broad variable emission with an irregular shape at a wavelength position close to that of the line \mbox{He\,{\sc ii}} 4686 was visible only (Fig.~\ref{HeII_06}). On July 9 it was much weaker compared to July 8, undergoing thus strong variation on a time-scale of about one day. At some times its long wavelengths side was partly absorbed by the high-velocity P~Cyg component of the line \mbox{He\,{\sc i}} 4713. We suppose that both components of the line \mbox{He\,{\sc ii}} 4686 contributed to this variable emission. 

The energy flux of the line \mbox{He\,{\sc ii}} 4686 was about 3.6 $\times$ 10$^{-11}$  erg\,cm$^{-2}$\,s$^{-1}$ in quiescence \citep{TTB08} and decreased to about 5 $\times$ 10$^{-12}$ erg\,cm$^{-2}$\,s$^{-1}$ at the time of the light maximum, which means that its emitting region was probably decreased and was related only to the close vicinity of the hot companion. 
At the time of the light maximum the line \mbox{He\,{\sc ii}} 4686 had a high negative velocity of less than $-$200\kms, which increased to about $-$20\kms\, in December (Table 5). This means that the motion of the emitting gas differs from the motion of the mass center of the system, whose velocity is $-$2\kms\, \citep{MK,f+00}. As in our previous work \citep{TTB08}, we suppose that the narrow component of the line \mbox{He\,{\sc ii}} 4686 is emitted in a region of a shock ionization where the outflowing material from the outbursting object meets the accretion disc and disc-like envelope. According to the gas dynamical modeling this region is heated mainly by propagation of a shock wave which begins to form about 70 days after the onset of the hot wind and reaches its maximal development about 20--30 days later \citep{Bis06}. This period of about 100 days is close to the typical time of the growth of the optical light during the outbursts of Z~And which means that the region of shock wave should form close to the light maximum and exist after it. Its temperature is higher than that of the surrounding medium and can reach 10$^6$ K. Some part of this region is occulted by the effective photosphere (the flat shell) of the outbursting object (Fig.~\ref{model2}). It is possible that gas particles in this region move radially outwards determining the width of the line. 

\begin{figure}[!tH]
\center
    \includegraphics[width=\linewidth]{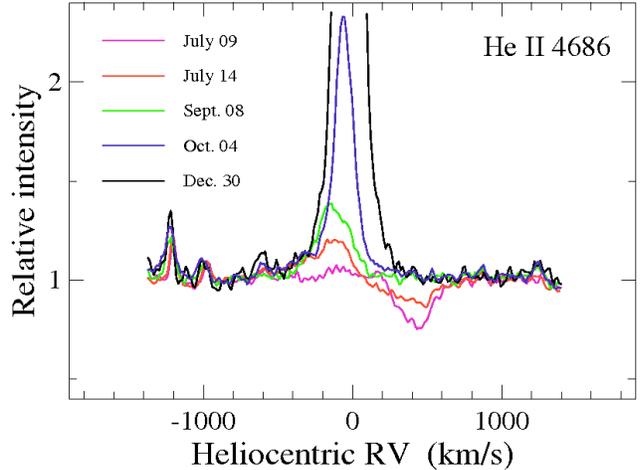}
\caption[The increase of the intensity of the \mbox{He\,{\sc ii}} 4686 narrow component and its gradual shift to the long wavelengths side]
				{The increase of the intensity of the \mbox{He\,{\sc ii}} 4686 narrow component and its gradual shift to the long wavelengths side with the decrease of the light.
				}
\label{HeIIN_06}
\end{figure}

After the light maximum the mass-loss rate of the compact object decreased (see the next section) which moved the level of the effective photosphere back to the star (Fig.~\ref{model2}). The size of this photosphere decreased and it occulted less and less the shock region. This determined both the increase of the intensity of the \mbox{He\,{\sc ii}} 4686 narrow component and its shift to the longer wavelengths (Fig.~\ref{HeIIN_06}) since more receding particles in the back part of its emitting region are seen.  

\subsubsection{\mbox{N\,{\sc iii}} and \mbox{C\,{\sc iii}} lines}

\begin{table*}[!tH]
 \centering
  \caption[The \mbox{N\,{\sc iii}} and \mbox{C\,{\sc iii}} lines data]{The \mbox{N\,{\sc iii}} and \mbox{C\,{\sc iii}} lines data.  RV and FWHM are in units \kms\, and $F$ -- in units $10^{-12}$ erg\,cm$^{-2}$\,s$^{-1}$.}
    \label{NIII,CIII_data}
  \begin{tabular}{@{}lrccrccrcc@{}}
\noalign{\smallskip}
  \hline\hline
    \noalign{\smallskip}
       Date & \multicolumn{3}{c}{\mbox{N\,{\sc iii}} $\lambda$ 4634} & \multicolumn{3}{c}{\mbox{N\,{\sc iii}} $\lambda$ 4641} & \multicolumn{3}{c}{\mbox{C\,{\sc iii}} $\lambda$ 4647} \\
       \cline{2-4} \cline{5-7} \cline{8-10} \noalign{\smallskip}
     				& RV & FWHM & $F$ & RV & FWHM & $F$ & RV & FWHM & $F$ \\
    \noalign{\smallskip}
  \hline
     \noalign{\smallskip}
     
Jul 8 & & & & $-$17 & ~~96 & 1.515 & $-$11 & ~~97 & 2.213 \\
Jul 9 & & & & & & & $-$25 & 108 & 1.206 \\
Jul 14 & & & & $-$47 & 136 & 2.160 & $-$30 & 113 & 2.088 \\
Jul 19 & & & & & & & $-$22 & 123 & 1.822 \\
Sep 7 & $-$17 & 100 & 1.019 & $-$43 & 171 & 3.012 & & &  \\
Sep 8 & $-$15 & ~~83 & 0.769 & & & & & & \\
Oct 3 & ~$-$3 & 107 & 1.365 & 6 & 158 & 3.465 & 12 & ~~89 & 2.043 \\
Oct 4 & $-$23 & 102 & 1.787 & $-$21 & 148 & 3.884 & $-$21 & 103 & 1.880 \\
Oct 31 & $-$13 & 113 & 2.535 & 0 & 107 & 4.986 & 2 & 100 & 2.701 \\
Dec 1 & ~$-$4 & ~~99 & 2.534 & 4 & ~~94 & 4.598 & 6 & ~~95 & 2.196 \\
Dec 2 & ~$-$4 & ~~94 & 2.315 & 1 & 102 & 4.648 & 2 & 105 & 2.432 \\
Dec 30 & $-$14 & ~~92 & 2.979 & $-$11 & ~~92 & 5.785 & $-$11 & ~~92 & 2.023 \\
    \noalign{\smallskip}
\hline
		\noalign{\smallskip}
\end{tabular}
\end{table*}

\begin{figure}[tH]
\center
    \includegraphics[width=\linewidth]{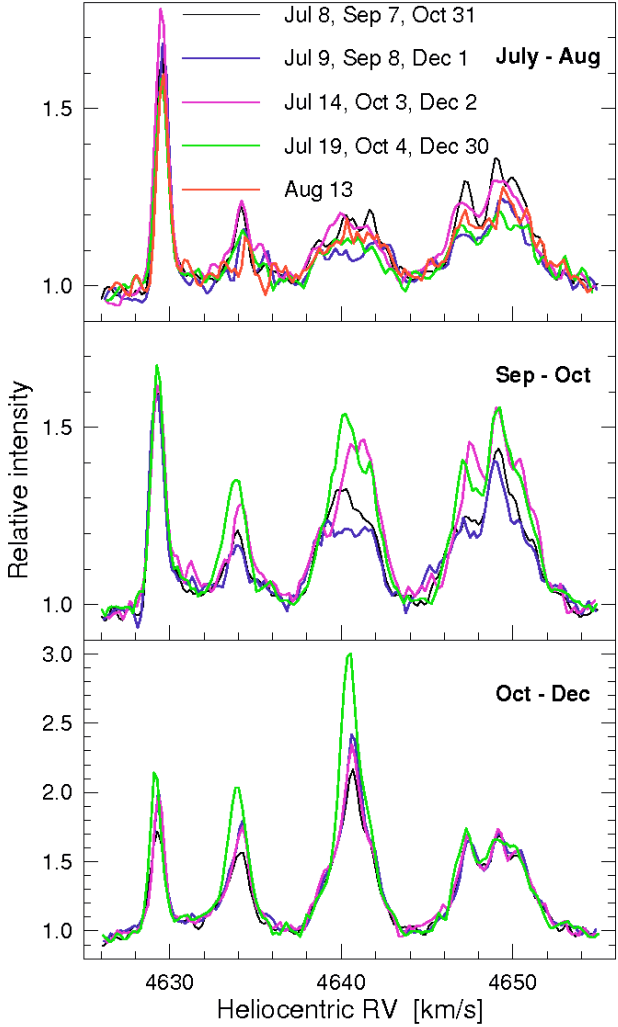}
\caption[The evolution of the profile of \mbox{N\,{\sc iii}} and \mbox{C\,{\sc iii}} lines]
				{The evolution of the profile of \mbox{N\,{\sc iii}} and \mbox{C\,{\sc iii}} lines: in July -- August (upper panel), September -- October (middle panel) and October -- December (lower panel)}
\label{NIII,CIII}
\end{figure}

We observed also some other lines of elements with high ionization degree -- \mbox{N\,{\sc iii}} 4634, \mbox{N\,{\sc iii}} 4641 and \mbox{C\,{\sc iii}}  4647. Their data are listed in Table~\ref{NIII,CIII_data}. The width of these lines increased and was well above its quiescent value of about 40\kms\, \citep{TTB08}. 

The line \mbox{N\,{\sc iii}} 4634 creates an unresolved blend with the line \mbox{Cr\,{\sc ii}} 4634. The relative intensity of the lines of high ionization degree strongly decreased at the time of maximal light and in July and August the nitrogen line was weaker than the line of \mbox{Cr\,{\sc ii}}. That is why in July and August the nitrogen line was not measured (Table~\ref{NIII,CIII_data}). 

\begin{figure}[!tH]
\center
    \includegraphics[width=.6\linewidth]{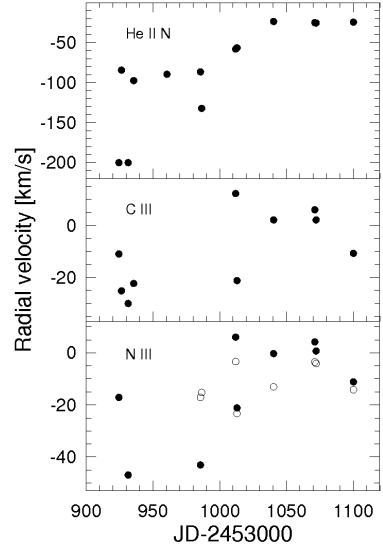}
\caption{The radial velocity of the lines of elements with high ionization degree. On the lower panel open circles mark the line \mbox{N\,{\sc iii}} 4634 and filled circles -- the line \mbox{N\,{\sc iii}} 4641. The uncertainty of the measurement reaches up to 5~\kms.}
\label{rad_vel_high_ion}
\end{figure}

During the outburst the line \mbox{N\,{\sc iii}} 4641 broadened and formed thus an unresolved blend with the lines of \mbox{O\,{\sc ii}} $\lambda$\,4639 and $\lambda$\,4642. The nitrogen lines were undergoing strong variation on a time-scale of about one day being on July 9 much weaker compared to July 8 like the line \mbox{He\,{\sc ii}} 4686. For this reason the line \mbox{N\,{\sc iii}} 4641 was very weak and was not measured on July 9. Moreover, it was weak and thus badly blended with the \mbox{O\,{\sc ii}} lines on July 19, Aug. 13 and Sept. 8 and was not measured too (Table~\ref{NIII,CIII_data}). Our data show that after the end of October the intensity of the \mbox{N\,{\sc iii}} lines gradually increased, without fluctuations (Table~\ref{NIII,CIII_data}, Fig.~\ref{NIII,CIII}). 

During the outburst the line \mbox{C\,{\sc iii}} 4647 broadened too and was blended with the lines \mbox{Fe\,{\sc ii}} 4649, \mbox{O\,{\sc ii}} 4649, \mbox{C\,{\sc iii}} 4650 and \mbox{O\,{\sc ii}} 4651. For this reason it was not measured on Aug. 13, Sept. 7 and Sept. 8. 

The radial velocity of the central narrow component of the line \mbox{He\,{\sc ii}} 4686 and the other lines of high ionization degree is shown in Fig.~\ref{rad_vel_high_ion}. The velocity of all lines increased. The data in Table~\ref{NIII,CIII_data} show that the flux of the nitrogen lines increases after the light maximum like the flux of the narrow \mbox{He\,{\sc ii}} component. This behaviour can be explained with the supposition that the nitrogen lines arise in the same region of shock wave whose unocculted part increases with the decrease of the mass-loss rate of the outbursting object.

\section{Discussion}

On one hand the star Z~And like other symbiotic systems (BF Cyg, Skopal et al. 2013; Hen 3-1341, Tomov et al. 2000) acquires satellite components of its spectral lines only during active phase when they are accompanied by a broad multicomponent strongly variable P~Cyg absorption. On the other hand, the presence of some lines of elements of high excitation and the obtained low effective temperature of the outbursting compact object during active phase can be explained only with the supposition that its pseudophotosphere is a disc-like (flat) shell placed in the orbital plane. This shell occults the compact object which leads to strong energy redistribution from UV range to longer wavelengths \citep{sk+06,sk+09}. These characteristics of the behaviour of Z~And can be easily explained in the framework of the model of collimated stellar wind. According to this model the wind of the outbursting compact object collides with the newly formed geometrically thick disc-shaped material which results in appearance of both an optically thick flat shell playing role of a pseudophotosphere and collimated outflow \citep{TTB14}. The area of the wind, which is projected onto the pseudophotosphere gives rise to the P~Cyg absorption. The receding part of the wind can give rise to the red wing of a broad emission, whose center should be at the radial velocity of the compact object (Fig. 1). 

The model of collimated stellar wind is related to the second stage of the evolution of the accreting compact object in the framework of the scenario proposed for interpretation of the line profiles of Z~And during its last active phase in 2000 -- 2013 \citep{TTB08,TTB14}. According to this scenario, in every binary system with parameters close to those of Z~And, some part of the ejected mass by the compact object during an outburst accretes again and can form extended disc-shaped structure which can collimate outflowing gas during the next outburst. In this way, an appearance of satellite components in the spectrum can be expected during active phase of every classical symbiotic star with parameters close to those of Z~And. A striking example is BF~Cyg \citep{sk+13}. After 2009 during activity, satellite components appeared in its spectrum and are observed up to now as it is seen in our unpublished spectra. Another system is Hen~3-1341, which had satellite components during its outbursts in 1999 \citep{tomata00} and 2012 \citep{TT13}. We suppose that satellite components can be expected during activity of other systems with orbital periods close to Z~And too, for example AG~Peg (after its new outbursts in 2015 it is treated as a classical symbiotic star \citep{ramsay16}), AX~Per and AG~Dra. 
Moreover, the idea of accretion of the part of the ejected matter sheds light to the question why there are more than one outburst taking place in some in classical symbiotic stars.

Another possibility to explain the line profiles of Z~And can be searched for in the framework of the commonly accepted model of highly collimated bipolar jets ejected by an accreting object. According to this model, a magnetic disc surrounding the compact object, exists whose magnetic field is responsible for the bipolar ejection. In this case, however, the next questions arise. Why do the satellite components appear only during an active phase? Is there connection between the stellar wind and the collimated outflow and what is it? These questions are difficult to answer. 

We calculated the mass-loss rate of the outbursting compact object using the energy fluxes of the broad emission component of the line \mbox{He\,{\sc ii}} $\lambda$\,4686 and the satellite components of \hb\ in the same way as in the work of Tomov et al. (2012). The rate was obtained the same like that based on the lines \ha\ and \hg\ considered in this work. That is why taking into account all four lines we conclude that the rate has decreased from 4--5 $\times$ 10$^{-7}$ (d/1.12kpc)$^{3/2}$ M$_{\sun}$\,yr$^{-1}$ at the time of the maximal light to 1--2 $\times$ 10$^{-7}$ (d/1.12kpc)$^{3/2}$ M$_{\sun}$\,yr$^{-1}$ at the end of December 2006. 

\section{Conclusion} 

We presented results of high-resolution observations in the range centered at $\lambda$\,4400 \AA\, and those  
of the lines \mbox{He\,{\sc ii}} $\lambda$\,4686, \hb\ and \mbox{He\,{\sc i}} $\lambda$\,6678 of the spectrum of the symbiotic binary Z~And carried out during 2006 eruption. The \hb\ had a multi-component profile consisting of an intense central narrow emission located around the reference wavelength, broad wings and high-velocity satellite emission features on both sides of the central emission at a velocity position of 1200--1300~km\,s$^{-1}$ indicating bipolar collimated ejection from the system. The broad emission wings and the satellite emissions were seen until the beginning of December 2006.

At almost all spectra the lines of \mbox{He\,{\sc i}} consisted of a narrow emission located close to the reference wavelength and a blueshifted absorption. At the time of the maximal light the absorption of the line \mbox{He\,{\sc i}} 4471 was multicomponent one occupying a broad velocity range like the line H$\gamma$. The most blueshifted component was at the velocity position of the satellite component. After that the broad absorption changed into a low velocity narrow feature, which was visible till the beginning of December 2006. 

In October and December the line \mbox{He\,{\sc ii}} 4686 had a two-component profile, consisting of a central narrow emission and a broad emission component with a low intensity. 

The observed lines are interpreted in the light of the model of collimated stellar wind \citep{TTB14}. This model provides a better interpretation of their profiles than the traditional model with a magnetic accretion disc. The scenario for evolution of the compact object proposed for interpretation of the line profiles of Z~And \citep{TTB08,TTB14} allows us to prognosticate appearance of satellite components of other classical symbiotic systems with close orbital periods and parameters. 

The mass-loss rate of the hot compact companion was found to decrease, from 4--5 $\times$ 10$^{-7}$ (d/1.12kpc)$^{3/2}$ M$_{\sun}$\,yr$^{-1}$ to about 1--2 $\times$ 10$^{-7}$ (d/1.12kpc)$^{3/2}$ M$_{\sun}$\,yr$^{-1}$ with the fading of the light after its maximum.  

\acknowledgments
The authors are thankful to the anonymous referee whose comments and suggestions led to improve the paper.
This work has been supported in part by the Basic Research Program of the Presidium of the Russian Academy 
of Sciences, Russian Foundation for Basic Research 
(projects 14-02-00215, 14-29-06059), and the Russian and Bulgarian Academies of Sciences through a collaborative program in Basic Space Research.
NAT and MTT gratefully acknowledge observing grant support from the Institute of Astronomy and Rozhen National Astronomical Observatory, Bulgarian Academy of Sciences.

\end{document}